\documentstyle[12pt]{article}

\newlength{\dinwidth}
\newlength{\dinmargin}
\begin{document}

\titlepage
\begin{flushright}
DTP/96/62  \\
August 1996 \\
\end{flushright}

\begin{center}
\vspace*{2cm}
{\Large \bf Forward $\pi^0$ trigger of the deep inelastic + jet probe of 
BFKL dynamics} \\
\vspace*{1cm}
J.\ Kwiecinski$^{a,b}$, S.\ C.\ Lang$^{a,b}$ and A.\ D.\
Martin$^b$ \\
\end{center}
\vspace*{0.5cm}

\begin{center}
$^a$ Department of Theoretical Physics, \\
H.\ Niewodniczanski Inst.\ of Nuclear Physics, \\
ul.\ Radzikowskiego 152, \\
31-342 Krakow, Poland. \\

\vspace*{0.5cm}
$^b$ Department of Physics, \\
University of Durham, \\
Durham, DH1 3LE, UK.
\end{center}

\vspace*{2cm}

\begin{abstract}
We predict the rate of deep inelastic scattering (DIS) events containing an 
identified forward $\pi^0$ that is expected in the experiments at the HERA 
electron-proton collider, in order to see if this process can be used as an 
indicator of the underlying small $x$ dynamics.  We determine the background 
due to deep inelastic events containing forward photons which are fragments of 
the forward jet.  We compare the DIS + $\pi^0$ cross section with that of the 
DIS + parent jet process.
\end{abstract}
\newpage

\section*{1. Introduction}

The behaviour of the proton structure function $F_2 (x, Q^2)$ at small $x$ 
reflects the behaviour of the gluon distribution, since the gluon is by far 
the dominant 
parton in this regime.  Perturbative QCD does not predict the absolute value of 
the parton distributions, but rather determines how they vary from a given input.  
If, for example, we are given initial distributions at some scale $Q_0^2$, then 
the DGLAP \cite{DGLAP} evolution equations enable us to determine the distributions 
at higher $Q^2$.  DGLAP evolution resums the leading $\alpha_S \ln (Q^2/Q_0^2)$ 
terms.  At sufficiently high electron-proton centre-of-mass energy $\sqrt{s}$ 
we encounter a second large variable\footnote{To be precise we define $x$ to be 
the Bjorken variable $x \equiv Q^2/2p.q$ where $Q^2 \equiv -q^2$, and $p$ and $q$ 
are the four momenta of the proton and virtual photon deep inelastic probe 
respectively.} $1/x \sim s/Q^2$, and we must resum the leading $\alpha_S \ln 
1/x$ contributions.  At leading order the resummation is accomplished by the 
BFKL equation \cite{BFKL} for the (unintegrated) gluon distribution.  The 
solution of the equation leads to a singular $x^{-\lambda}$ small $x$ behaviour 
of the gluon distribution, where $\lambda = (3 \alpha_S/\pi) 4 \ln 2$ for fixed 
$\alpha_S$ and $\lambda \simeq 0.5$ if a reasonable prescription for the running 
of $\alpha_S$ is assumed and for the treatment of the infrared region \cite{AKMS}.  
The $x^{-\lambda}$ behaviour of the BFKL gluon feeds through, via the 
$k_T$-factorization theorem \cite{KTFAC}, into the small $x$ behaviour of the 
structure function $F_2$.  Of course in practice we should not expect such a 
dramatic growth with decreasing $x$, since subleading effects are expected to 
suppress the effective value of $\lambda$ in the HERA domain \cite{SL}.

However, it is difficult to identify the presence of the $\alpha_S \ln 1/x$ terms 
in the measurements of $F_2$ at HERA even though the data do show a steep rise 
with decreasing $x$.  In fact the rise in the latest precise H1 and ZEUS 
measurements \cite{HZ} can be well described by next-to-leading order
DGLAP evolution down to $Q^2 
\sim 2$ GeV$^2$ and $x \sim 10^{-5}$.  The problem in identifying the underlying 
small $x$ dynamics is due to the parametric freedom that we have in specifying 
the initial parton distributions.  For instance for a non-singular gluon input 
we can increase the steepness of $F_2$ with decreasing $x$ by simply reducing 
$Q_0^2$ and increasing the DGLAP evolution length, $\ln (Q^2/Q_0^2)$.  
Alternatively, we could use (as in the global parton analyses \cite{MRS,CTEQ}) 
a singular input form $xg (x, Q_0^2) \sim x^{-\lambda}$, with $\lambda$ chosen 
to fit the data.  We conclude that it is difficult to isolate $\ln 1/x$ effects 
from measurements of $F_2$.  The observable $F_2$ is too inclusive.  Rather, we 
should explore properties of the final state in deep inelastic scattering (DIS).

The classic way \cite{M} to probe the small $x$ behaviour of QCD, which 
avoids the problem of assuming input parton distributions, is to study deep 
inelastic $(x, Q^2)$ events which contain an identified forward jet $(x_j, 
k_{jT}^2)$, see Fig.\ 1(a).  According to BFKL dynamics the 
differential structure function for DIS + jet events has the following
small $(x/x_{j})$ behaviour
\begin{equation}
\frac{ \partial F_2} {\partial \left(\ln 1/x_{j}\right) \partial 
k_{jT}^2} = C \alpha_s (k_{jT}^2) x_{j} \left[g+ \frac{4}{9} \left(
q + \bar q \right)\right] 
\left(\frac{x}{x_{j}}\right)^{-\lambda} ,
\label{eq:a1}
\end{equation}
where the normalisation coefficient $C$ is given in refs.\ 
\cite{T,BD,KMS1,KMS2}.
The parton distributions $g$, $q$ and $\bar q$ are to be evaluated
at $(x_{j},k_{jT}^2)$.  The relevant kinematic region is where
\begin{itemize}
\item[(i)] the jet transverse momentum satisfies $k_{jT}^2 \simeq Q^2$ 
so as to neutralize the DGLAP evolution, and is sufficiently large so as to 
suppress diffusion into the infrared region when we solve the BFKL equation 
at decreasing values of $x/x_j$;
\item[(ii)] the jet longitudinal momentum $x_j p$ is as large as is experimentally 
feasible (and $x$ is as small as possible) so as to be able to probe the region of 
small $x/x_j$.
\end{itemize}
\noindent For these values of $x_j$ the parton distributions entering (\ref{eq:a1}) 
are well known from the global parton analyses and so the observation of DIS + 
jet events offers the opportunity to expose BFKL-type small $x$ dynamics free 
from the ambiguities associated with the choice of the non-perturbative parton 
input.  In other words we are studying small $x$ dynamics by deep inelastic 
scattering off a known parton, rather than off the proton.  Experimentally, 
however, the clean identification and kinematic measurement
of a forward jet proves to be difficult since we require it to be as close to 
the proton remnants as possible, that is $x_j$ as large as possible.  
Nevertheless experimental studies have been attempted and lead to encouraging 
results \cite{H1}.

Here we use the improved knowledge of the fragmentation functions
to propose that the forward jet is identified through the measurement of a 
single energetic decay product. As it turns out the $\pi^0$ is the hadron
which can be identified in the most forward direction in the
detectors at HERA.  We use the BFKL formalism to predict the DIS + forward 
$\pi^0$ rate.  The rate will, of course, be suppressed in comparison with the 
DIS + forward jet rate and it is an experimental issue to see if the advantages 
of single particle detection as compared to identification of the (parent) jet 
can compensate for the loss of signal.

The outline of the paper is as follows. In section 2 we present the
QCD formalism required to calculate the cross section for the
deep inelastic + forward $\pi^{0}$ process. Then in section 3 we discuss
the experimental cuts which we impose and give our numerical
predictions for the DIS + $\pi^{0}$ cross section. Section 4
contains a discussion.

\section*{2. The DIS + forward $\pi^0$ cross section}

First we recall the derivation of the cross section for the deep inelastic + 
jet process depicted in Fig.\ 1(a), which also shows the variables used.  The 
differential cross section is given by \cite{KMS2}
\begin{equation}
\frac{\partial \sigma_{j}}{\partial x \partial Q^{2}} = \int dx_{j}
\int dk_{jT}^{2} \frac{4 \pi \alpha ^{2}}{xQ^{4}} \left[
\left(1-y \right) \frac{\partial F_{2}}{\partial x_{j} \partial
k_{jT}^{2}} + \frac{1}{2} y^{2} \frac{\partial F_{T}}
{\partial x_{j} \partial k_{jT}^{2}} \right] 
\label{eq:a2}
\end{equation}
where the differential structure functions have the following
leading small $x/x_{j}$ form
\begin{equation}
\frac{\partial^2 F_{i}}{\partial x_{j} \partial k_{jT}^{2}} =
\frac{3 \alpha_S \left(k_{jT}^{2} \right) }{\pi k_{jT}^{4}}
\sum_{a} f_a \left(x_{j},k_{jT}^{2} \right) \Phi_i
\left(\frac{x}{x_{j}},k_{jT}^{2},Q^{2} \right)
\label{eq:a3}
\end{equation}
for $i=T,L$. Assuming $t$-channel pole dominance the sum over 
the parton distributions is given by
\begin{equation}
\sum_{a} f_{a} = g+\frac{4}{9} \left(q+ \bar q \right).
\label{eq:a4}
\end{equation}
Recall that these parton distributions
are to be evaluated at $(x_{j},k_{jT}^2)$ where 
they are well-known from the global analyses, so there are no ambiguities
arising from a non-perturbative input.

The functions $\Phi_i (x/x_j, k_{jT}^{2}, Q^2)$
in (\ref{eq:a3}) describe the virtual $\gamma$ + virtual gluon fusion process 
including the ladder formed from the gluon chain of Fig.\ 1(a).  They can be 
obtained by solving the BFKL equation
\begin{eqnarray}
\Phi_i (z, k_T^2, Q^2) & = & \Phi_i^{(0)} (z, k_T^2, Q^2) \: +
\nonumber \\
& + & \frac{3 \alpha_S}{\pi} \: k_T^2 \: \int_z^1 \:
\frac{dz^\prime}{z^\prime} \: \int_0^\infty \:
\frac{dk_T^{\prime 2}}{k_T^{\prime 2}} \; \left [ \frac{\Phi_i
(z^\prime, k_T^{\prime 2}, Q^2) \: - \: \Phi_i (z^\prime, k_T^2,
Q^2)}{| k_T^{\prime 2} \: - \: k_T^2 |} \: + \: \frac{\Phi_i
(z^\prime, k_T^2, Q^2)}{\sqrt{4k_T^{\prime 4} \: + \: k_T^4}} \right ]. \nonumber \\
& & 
\label{eq:a5}
\end{eqnarray}
The inhomogeneous or driving terms $\Phi_{i}^{(0)}$ correspond
to the sum of the quark box and crossed-box contributions.
For small $z$ we have
\begin{equation}
\Phi_i^{(0)} (z, k_T^2, Q^2) \; \approx \; \Phi_i^{(0)} (z = 0,
k_T^2, Q^2) \; \equiv \; \Phi_i^{(0)} (k_T^2, Q^2).
\label{eq:a6}
\end{equation}
We evaluate the $\Phi_i^{(0)}$ by expanding the four momentum in terms of the 
basic light-like four momenta $p$ and $q^\prime \equiv q + xp$.  For example, 
the quark momentum ${\kappa}$ in the box (see Fig.\ 1(a)) has the Sudakov 
decomposition
$$
\kappa \; = \; \alpha p \: - \: \beta q^\prime \: + \: \mbox{\boldmath $\kappa$}_T.
$$
We carry out the integration over the box diagrams, subject to the quark 
mass-shell constraints, and find
\begin{eqnarray}
\Phi_T^{(0)} (k_T^2, Q^2) & = & 2 \: \sum_q \: e_q^2 \:
\frac{Q^2}{4 \pi^2} \: \alpha_S \: \int_0^1 \: d \beta \: \int \:
d^2 \kappa_T \left [\beta^2 \: + \: (1 - \beta)^2 \right ] \left (
\frac{\kappa_T^2}{D_1^2} \: - \: \frac{\mbox{\boldmath $\kappa$}_T .
(\mbox{\boldmath $\kappa$}_T - \mbox{\boldmath $k$}_T)}{D_1 D_2}
\right ) \nonumber \\
& & \nonumber \\
\Phi_L^{(0)} (k_T^2, Q^2) & = & 2 \sum_q \: e_q^2 \;
\frac{Q^4}{\pi^2} \: \alpha_S \: \int_0^1 \: d \beta \: \int \:
d^2 \kappa_T \: \beta^2 (1 - \beta)^2 \; \left ( \frac{1}{D_1^2} \:
- \: \frac{1}{D_1 D_2} \right ).
\label{eq:a7}
\end{eqnarray}
where the denominators $D_{i}$ are of the form
\begin{eqnarray}
D_1 & = & \kappa_T^2 \: + \: \beta (1 - \beta) \: Q^2 \nonumber \\
& & \\
D_2 & = & (\mbox{\boldmath $\kappa$}_T - \mbox{\boldmath $k$}_T)^2 \:
+ \: \beta (1 - \beta) \: Q^2, \nonumber
\label{eq:a8}
\end{eqnarray}
assuming massless quarks.

If the QCD coupling $\alpha_S$ is fixed we can solve the BFKL
equation (5) and obtain an analytic expression for the leading
small $z$ behaviour of the solution. Omitting the Gaussian diffusion
factor in $\ln \left(k_T^{2}/Q^{2} \right)$ we find
\begin{eqnarray}
\Phi_T (z, k_T^2, Q^2) & = & \frac{9 \pi^2}{512} \: \frac{2 \sum
e_q^2 \alpha_S^{\frac{1}{2}}}{\sqrt{21 \zeta (3)/2}} \; (k_T^2
Q^2)^{\frac{1}{2}} \; \frac{z^{- \alpha_P + 1}}{\sqrt{\ln (1/z)}}
\; \left [ 1 \: + \: {\cal O} \: \left ( \frac{1}{\ln (1/z)}
\right ) \right ] \nonumber \\
& & \\
\Phi_L (z, k_T^2, Q^2) & = & \frac{2}{9} \: \Phi_T (z, k_T^2, Q^2)
\nonumber
\label{eq:a9}
\end{eqnarray}
where the Riemann zeta function $\zeta (3) = 1.202$ and the BFKL
intercept
\begin{equation}
\alpha_P \: - \: 1 \; = \; \frac{12 \alpha_S}{\pi} \; \ln 2.
\label{eq:a10}
\end{equation}

Here, however, we follow the approach of \cite{KMS1} and allow the
coupling $\alpha_{S}$ to run. This means that we must numerically
solve the BFKL equations for
\begin{equation}
H_i (z, k_T^2, Q^2) \; \equiv \; \frac{3 \alpha_S (k_T^2)}{\pi} \; \Phi_i
(z, k_T^2, Q^2).
\label{eq:a11}
\end{equation}
We use the differential form of the equations,
\begin{equation}
\frac{\partial H_i (z, k_T^2, Q^2)}{\partial \ln (1/z)} \; = \;
\frac{3 \alpha_S (k_T^2)}{\pi} \: k_T^2 \: \int_{k_0^2}^\infty \:
\frac{dk_T^{\prime 2}}{k_T^{\prime 2}} \; \left [ \frac{H_i (z,
k_T^{\prime 2}, Q^2) - H_i (z, k_T^2, Q^2)}{| k_T^{\prime 2} - k_T^2 |}
\: + \: \frac{H_i (z, k_T^2, Q^2)}{\sqrt{4k_T^{\prime 4} \: + \:
k_T^4}} \right ]
\label{eq:a12}
\end{equation}
subject to the boundary conditions
\begin{equation}
H_i (z = z_0, k_T^2, Q^2) \; = \; H_i^{(0)} (k_T^2, Q^2).
\label{eq:a13}
\end{equation}
For the lower limit on the transverse momentum integration we
choose $k_0^2 = 1$ GeV$^2$.  We start from the \lq\lq box" expressions, 
(\ref{eq:a7}), for $H_i^{(0)}$ at $z_0 = 0.1$ and solve (\ref{eq:a12}) to obtain 
$H_i$ (and $\Phi_i$) for $z < z_0$.  In this way we predict the cross section for 
DIS + jet production from equations (\ref{eq:a2}) and (\ref{eq:a3}).

Next let us consider the process where the forward jet fragments into
$\pi^0$'s as shown schematically in Fig.\ 1(b). We are looking at the
case where the $\pi^0$ is collinear with the parent quark jet.
This means that if the $\pi^0$ carries a fraction $x_{\pi}$ of the 
proton's longitudinal momentum, then it carries a 
fraction $z=x_\pi/x_j$ ($0 \leq z \leq 1$)
of the parent jet's longitudinal momentum
and its transverse momentum $k_{\pi T}=z k_{j T}$. In order to
calculate the cross section for DIS + $\pi^0$ production we have
to convolute the DIS + jet cross section with the $\pi^0$
fragmentation functions. We obtain
\begin{eqnarray}
\frac{\partial \sigma_{\pi}}{\partial x_{\pi} \partial k_{\pi T}} = 
\int_{x_{\pi}}^{1} dz \int dx_{j} \int dk_{j T}^{2}
\left[ \frac{\partial \sigma_{g}}{\partial x_{j} \partial k_{jT}^{2}} 
D_{g}^{\pi^0} \left( z,k_{\pi T}^{2} \right) + \sum_{q}
\left( \frac{\partial \sigma_{q}}{\partial x_{j} \partial k_{jT}^{2}}
D_{q}^{\pi^0} \left( z,k_{\pi T}^{2} \right) \right) \right] 
\times \nonumber \\
\times \delta \left(x_{\pi} - zx_{j} \right)
\delta \left(k_{\pi T} - zk_{jT} \right) 
\label{eq:a14}
\end{eqnarray}
where the sum over $q$ runs over all quark and antiquark flavours.
The partonic differential cross sections can be obtained from 
(\ref{eq:a2}) and (\ref{eq:a3})
by substituting for the sum over the parton distributions
$\sum_{a} f_{a}$ either the gluon distribution $g$ or the quark
or antiquark distribution $\frac{4}{9} q$ or $\frac{4}{9} \bar q$
respectively.  In analogy to choosing $z_{0}=0.1$ in (\ref{eq:a13}) we 
impose the constraint $x/x_{\pi}<0.1$, i.e. $x/x_{j}<0.1$ since 
$x_{\pi}<x_{j}$, on the $x_{j}$ integration here.
The functions $D_{g}^{\pi^0} ( z,k_{\pi T}^{2})$ and 
$D_{q}^{\pi^0} ( z,k_{\pi T}^{2})$ in (\ref{eq:a14}) give the 
probability that a
gluon or quark jet fragments into a $\pi^{0}$ carrying a fraction
$z$ of the parent jet's momentum. We assume that these fragmentation functions
satisfy leading order DGLAP evolution equations. Note that SU(2) 
isospin symmetry implies that 
\begin{equation}
D_{i}^{\pi^0} ( z,k_{\pi T}^{2}) \; = \; \frac{1}{2} \left(D_{i}^{\pi^+} 
( z,k_{\pi T}^{2}) \: + \: D_{i}^{\pi^-} ( z,k_{\pi T}^{2})
\right)
\label{eq:a15}
\end{equation} 
for all partons $i =q,g$. Therefore (\ref{eq:a14}) 
describes the average of the cross sections for $\pi^+$ and $\pi^-$
production. We use the parametrizations
of the leading order charged pion fragmentation functions obtained by 
Binnewies et al.\ \cite{CH}; their analysis treated light, s, c, and
b quarks independently for the first time.  

\section*{3. Predictions for the DIS + $\pi^0$ cross section}
 
We use (\ref{eq:a14}) to calculate the event rate for deep inelastic scattering 
in which, in addition to the scattered electron, the $\pi^0$ is measured in the 
final state.  To ensure that the $\pi^0$ is really a fragment of the forward 
jet (and does not come from the quark-antiquark pair which form the quark box) 
we require the $\pi^0$ to be emitted in the forward hemisphere in the virtual 
photon-proton centre-of-mass frame.  If we express the pion four momentum in 
terms of Sudakov variables
\begin{equation}
k_\pi \; = \; x_\pi p \: + \: \beta_\pi q^\prime \: + \: 
\mbox{\boldmath $k$}_{\pi T}
\label{eq:a16}
\end{equation}
then the forward hemisphere requirement is
\begin{equation}
x_\pi \; > \; \beta_\pi.
\label{eq:a17}
\end{equation}
Since the outgoing pion satisfies the on-mass-shell condition 
$k_\pi^2 = m_\pi^2 \approx 0$ we have
\begin{equation}
\beta_\pi \; = \; \frac{x}{x_\pi} \: \frac{k_{\pi T}^2}{Q^2}.
\label{eq:a18}
\end{equation}
Then (\ref{eq:a17}) gives
\begin{equation}
x_j \; > \; x_\pi \; > \; \sqrt{x k_{\pi T}^2/Q^2}.
\label{eq:a19}
\end{equation}
We thus have an implicit lower limit on the $x_j$ integration in (\ref{eq:a14}), 
which is generally stronger than the condition $x_j > 10 x$ imposed on the 
solution of the BFKL equation.

Another problem to be taken into account is that at HERA pions can
only be detected if they are emitted at a large enough angle to the
proton beam. This also ensures that there is no contamination
from pions produced in the proton remnant. We require
\begin{equation}
\theta_{\pi p} > \theta_{0}.
\label{eq:a20}
\end{equation}
In Fig.\ 2 we show the relation between the pion kinematic variables
for different choices of the minimum angle $\theta_{0}$ defined in the 
HERA frame.  We find that
pions with large longitudinal momentum fraction $x_{\pi}$ are only
emitted at small angles $\theta_{\pi p}$. To reach larger $x_{\pi}$ 
for a given $\theta_{\pi p}$ we can measure pions with larger
$k_{\pi T}^{2}$ but at a depleted event rate. In the same figure
we also plot the boundary given by the hemisphere cut, (\ref{eq:a17}),
for $x = 6 \times 10^{-4}$ and $Q^2 = 20$ GeV$^2$, which acts 
as a lower limit on the allowed kinematic region. We will use 
$\theta_{0}=5^\circ$ for the main presentation of our results (although 
later we compare the predictions with those obtained with $\theta_0 = 
7^\circ$).

Now we are in the position to give numerical predictions for the 
cross section for the DIS + $\pi^{0}$ production using 
(\ref{eq:a14}) and implementing the cuts that we just discussed.
Recall that it follows from (\ref{eq:a15}) that the cross 
section for $\pi^0$ production equals the average of the cross
sections for $\pi^+$ and $\pi^-$ production. Therefore the 
results we will show in the following multiplied by a factor
of 2 will be valid for charged pion production. Throughout
the analysis we assumed three flavours of massless quarks.
In Fig.\ 3 we plot the $x$ dependence of this cross section 
integrated over bins of size $\Delta x=2.10^{-4}$ and
$\Delta Q^{2}=10$ GeV$^2$ for three different $Q^{2}$ bins,
namely 20-30, 30-40, 40-50 GeV$^2$. Here we required that
$x_{\pi}>0.05$ and $3< k_{\pi T} < 10$ GeV and used the
fragmentation functions \cite{CH} at scale $k_{\pi T}^{2}$. We compare the
results obtained when BFKL small $x$ resummation is included
with the case when gluon radiation is neglected. In the first
case the strong $x$ dependence of the cross section is driven
by the small $z$ behaviour of the $\Phi_{i}$ and therefore
there is a strong enhanced increase with decreasing $x$. 
For example, if we compare
the cross section for $x \approx 5 \times 10^{-4}$ in the two cases, 
we find that the results are about a factor 7 larger when the BKFL 
resummation is included than when it is neglected. This 
enhancement is the signature of BFKL soft gluon resummation.
In fact the BFKL behaviour should be identified via the
shape in $x$ rather than the value of the cross section, since
the latter is subject to normalisation uncertainties \cite{KMS1}.
In Fig.\ 4 we show the cross section (in fb), for the same cuts as in 
Fig.\ 3, in various bins in $x$ and $Q^{2}$ which are accessible at HERA.
We find that the cross section drops off very rapidly with $Q^2$
which means that we can reach the highest values for the bins with  
$10<Q^{2}<20$ GeV$^2$ and $x$ very small.

Of course the DIS + $\pi^{0}$ cross section depends on the values
chosen for the cuts. In table 1 we show the effect of changing
the limits on the $k_{\pi T}$ integration. Since the cross section
decreases with increasing $k_{\pi T}$ it is more sensitive to
the lower limit on the $k_{\pi T}$ integration
than to the upper limit.

\begin{table}[htb]
\caption{The DIS + $\pi^{0}$ cross section in the bin $
10^{-3} < x < 1.2 \times 10^{-3}$, $20 < Q^2 < 30$ GeV$^2$ as
calculated in Fig.\ 4, but for different choices of
the limits of the integration over the transverse momentum of the $\pi^{0}$.}
\begin{center}
\begin{tabular}{|c|c|c|} \hline
$k_{\pi T,min}$ [GeV] & $k_{\pi T,max}$ [GeV] & $\sigma$ [pb] \\
\hline
3 & 8 & 0.23 \\
3 & 10 & 0.26 \\
5 & 10 & 0.18 \\ \hline
\end{tabular}
\end{center}
\end{table}

\newpage

To obtain the results shown in Figs.\ 3 and 4 we used the pion
fragmentation functions at scale $k_{\pi T}^{2}$. In  table 2 we
show the cross section for the deep inelastic + $\pi^{0}$ process in 
the bin $10^{-3} < x < 1.2 \times 10^{-3}$, $20 < Q^{2} < 30$ GeV$^2$
calculated imposing the same constraints and including BFKL soft gluon 
resummation but evaluating the pion fragmentation functions at the scales 
$\frac{1}{2} k_{\pi T}^{2}$, $k_{\pi T}^{2}$ and $2k_{\pi T}^{2}$.  The 
values demonstrate the scale ambiguity in the prediction of the cross 
section.

\begin{table}[htb]
\caption{The DIS + $\pi^{0}$ cross section in the bin $
10^{-3} < x < 1.2 \times 10^{-3}$, $20 < Q^2 < 30$ GeV$^2$ 
calculated imposing the same cuts as for Fig.\ 4, but 
evaluating the fragmentation functions at the three different
scales $\frac{1}{2} k_{\pi T}^{2}$, $k_{\pi T}^{2}$ and $2k_{\pi T}^{2}$.}
\begin{center}
\begin{tabular}{|c|c|} \hline
fragmentation scale &  $\sigma$ [pb] \\
\hline
$\frac{1}{2} k_{\pi T}^{2}$ & 0.31 \\
$k_{\pi T}^{2}$ & 0.26 \\
$2k_{\pi T}^{2}$ & 0.23 \\ \hline
\end{tabular}
\end{center}
\end{table}

Since $\pi^{0}$'s are measured through their decay into two photons
there is a background from events in which the parent jet fragments
into a photon which is emitted collinearly, see Fig.\ 5. In 
analogy to (\ref{eq:a14}) the corresponding cross section is 
given by
\begin{eqnarray}
\frac{\partial \sigma_{\gamma}}{\partial x_{\gamma} \partial k_{\gamma T}} = 
\int_{x_{\gamma}}^{1} dz \int dx_{j} \int dk_{j T}^{2}
\left[ \frac{\partial \sigma_{g}}{\partial x_{j} \partial k_{jT}^{2}} 
D_{g}^{\gamma} \left( z,k_{\gamma T}^{2} \right) + \sum_{q}
\left( \frac{\partial \sigma_{q}}{\partial x_{j} \partial k_{jT}^{2}}
D_{q}^{\gamma} \left( z,k_{\gamma T}^{2} \right) \right) \right] 
\times \nonumber \\
\times \delta \left(x_{\gamma} - zx_{j} \right)
\delta \left(k_{\gamma T} - zk_{jT} \right) .
\label{eq:a21}
\end{eqnarray}
We estimated this background using the fragmentation functions of
Owens \cite{OW} and found that it is 1-2 \% of the cross
section for $\pi^{0}$ production. 

Considering the smallness of the background from photons which are fragments
of the forward jet, a comment on the errors on the calculation of
the cross section for pion production is due here.  From the numerical point 
of view there is an error from the Monte-Carlo 
integration used to evaluate (\ref{eq:a14}) which is of the order of 5 \%. 
To our knowledge the errors on the pion fragmentation functions
are of the order of a few percent for quarks and 30 - 40 \% for
gluons \cite{B}. Since the dominant contribution to the cross
section for pion production comes from the fragmentation of gluons
we expect these errors on the fragmentation functions to result
in an error of at most 25 \% on the cross section. 
The parametrizations of the fragmentation
functions describe the DGLAP evolution correctly up to 10 \% \cite{CH}.
We found that our results are more sensitive to the normalisation
of the fragmentation functions than to their shape.

Finally let us compare our predictions for DIS + forward $\pi^0$ production 
as shown in Fig.~4, with the corresponding cross sections for the DIS + forward 
jet events, the process originally proposed by Mueller as the probe of small 
$x$ dynamics.  In order to quantify the suppression due to the fragmentation of 
the jet into the $\pi^0$ we integrate the DIS + jet differential structure 
functions given in (\ref{eq:a3}) over the same domains of $x_j$ and $k_{jT}^2$ 
that we used for $x_\pi$ and $k_{\pi T}^2$ for the DIS + $\pi^0$ predictions.  
To be precise we integrate over the region $3 < k_{jT} < 10$ GeV and $\theta_{jp} 
> 5^\circ$ with a hemisphere cut for the jet in analogy to (\ref{eq:a17}), that 
is $x_j > \beta_j$.  The upper and lower numbers in Fig.~6 compare the DIS + 
$\pi^0$ with the DIS + jet cross section in the various bins of $x$ and $Q^2$.  
We see that the fragmentation of the forward jet into a $\pi^0$ meson costs a 
factor of about 40 in the suppression in event rate.  Whether this loss of 
event rate is compensated by the advantage of identifying a forward $\pi^0$ 
as compared to a jet (adjacent to the proton remnants) is an experimental 
question.  Table 3 offers a guide to the possible gain using the $\pi^0$ 
signal.  For instance if we were able to identify $\pi^0$ mesons down to 
$5^\circ$ in angle and 5 GeV in $k_T$ with the same accuracy as jets down 
to $7^\circ$ in angle and 7 GeV in $k_T$ then we would gain back a factor 
of 4 \footnote{Since $x_\pi = z x_j$ with $\langle z \rangle \sim 0.4$, our choice of 
cut-off 
on $x_\pi$, that is $x_\pi > 0.05$, is probably too conservative \cite{MK}. 
If we were
to take $x_\pi > 0.02$ then the $\pi^0$ production rates are all increased.
For example if we choose $Q^2 \sim 25$ GeV$^2$ then for $x = 5.10^{-4}$ and 
$10^{-3}$ the results are enhanced by additional factors of 6 and 5 respectively.
Also the recent DIS + jet data \cite{JET} lie about a factor of 1.7 above 
the values that would be obtained from our choice of input, so the
DIS + $\pi^0$ rates are expected to be further enhanced by such a factor.}. 
Moreover if we were to add in the DIS + forward $\pi^\pm$ signal then 
we gain an extra factor of 3. Table 3 also shows that in the HERA regime, 
where we need to take $x_j$ sufficiently large (say $x_j > 0.05$) to make
$x/x_j$ small, the low $k_T$ events are kinematically forbidden by the
cuts. For example for $\theta_{0}=7^\circ$ we find that $k_T > 5.0$ GeV, while
for $\theta_{0}=5^\circ$ we have $k_T > 3.6$ GeV.

\begin{table}[htb]
\caption{The DIS + $\pi^{0}$ and DIS + jet cross sections in the bin 
$10^{-3} < x < 1.2 \times 10^{-3}$, $20 < Q^2 < 30$ GeV$^2$ as
calculated for Fig.~6, but integrated over domains with different choices of 
the minimum angle $\theta_0$ between
the proton and the $\pi^0$ or the jet, and of the minimum transverse
momentum $k_{T, min}$ of the $\pi^0$ or the jet.}
\begin{center}
\begin{tabular}{|c|c|c|c|} \hline
$k_{T,min}$ [GeV] & $\theta_{0}$ & $\sigma_{\pi^{0}}$ [pb]& 
$\sigma_{j}$ [pb] \\
\hline
3 & $5^\circ$ & 0.26 & 10.3 \\
3.5 & $5^\circ$ & 0.26 & 10.3 \\
5 & $5^\circ$ & 0.18 & 8.0 \\
7 & $5^\circ$ & 0.07 & 3.7 \\ \hline
3.5 & $7^\circ$ & 0.08 & 3.4 \\
5 & $7^\circ$ & 0.08 & 3.4 \\
7 & $7^\circ$ & 0.04 & 2.0 \\ \hline
\end{tabular}
\end{center}
\end{table}

\section*{4. Conclusion}

In principle, the DIS + jet measurement should be an excellent way
of identifying the BFKL soft gluon resummation effects at HERA. It
turns out, however, that it is experimentally quite difficult to
measure a forward jet so close to the proton remnants. We 
therefore suggested studying the fragmentation of this forward
jet into a single energetic decay product, the $\pi^{0}$.
This should be easier to measure.  (The DIS + $\pi^0$ signal can, 
of course, be supplemented by also observing jet fragmentation into 
$\pi^\pm$ mesons).  We found that when we include BFKL dynamics in the 
calculation of the cross section it leads to the characteristic steep rise 
with decreasing $x$.  The disadvantage of using 
the DIS + $\pi$ process is that the event rate is lower than for DIS + 
forward jet.  We quantified the suppression which arises from this jet 
to $\pi$ fragmentation (see also footnote 2).  
It is an experimental question as to whether the 
loss of event rate can be compensated by the more forward domain accessible 
for $\pi$ detection and the more accurate measurement of the kinematic 
variables possible for pions as opposed to jets.  We presented sample results 
for different acceptance cuts to help provide an answer.

Since $\pi^0$'s are measured via the 
two photon decay, there is a background to the deep-inelastic + $\pi^{0}$ 
measurement from events in which the
parent jet fragments into a photon which is being emitted
collinearly to the jet. We found that this background is
about 1-2 \%.  We conclude that deep-inelastic + pion events
should be a good way of probing small $x$ dynamics at HERA.

\section*{Acknowledgements}

We thank P.\ Sutton for valuable help, and J.\ Binnewies,
B.\ Kniehl and S.\ Rolli for discussion.
J.K.\ would like to thanks the Department of Physics and Grey College 
of the University of Durham for their warm hospitality.
S.C.L.\ thanks the UK Engineering and Physical Sciences Research Council 
for a Studentship and the H.\ Niewodniczanski Institute of Nuclear 
Physics in Krak\'{o}w for warm hospitality.  
This work has been supported in part by the Polish State Committee
for Scientific Research grant 2 P03B 231 08, by Stiftung f\"{u}r 
Deutsch-Polnische-Zusammenarbeit, project 1522/94/LN and by the EU 
under contracts no. CHRX-CT92-0004/CT93-357.\\

\newpage
\section*{Figure Captions}
\begin{itemize}
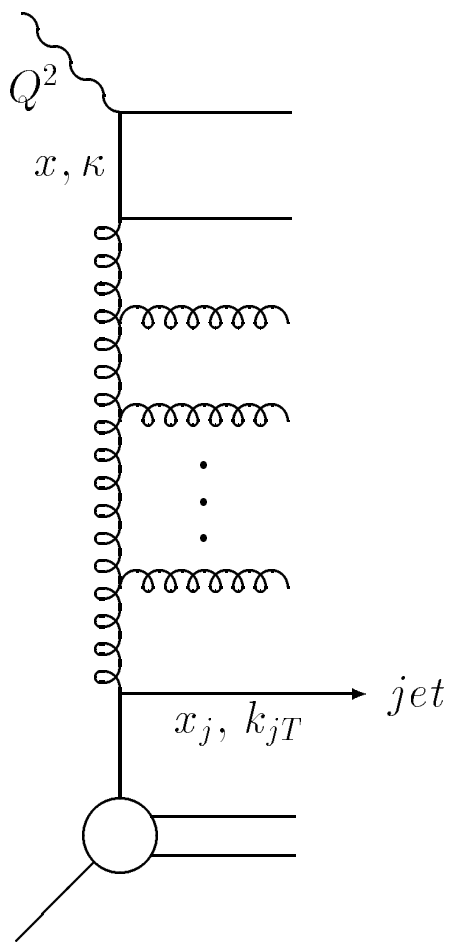
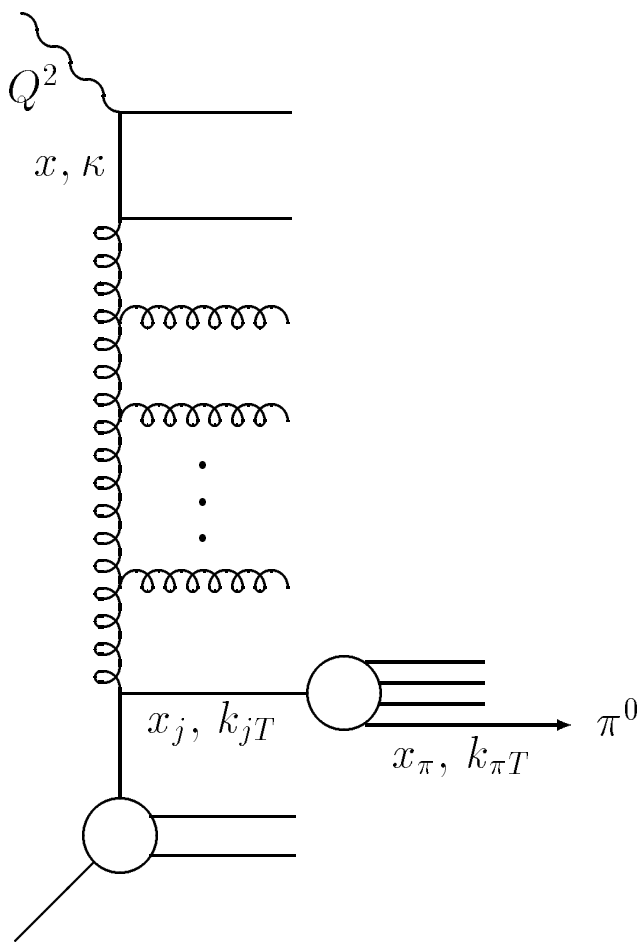
\item[Fig.\ 1] Diagrammatic representation of (a) a
deep inelastic + forward jet event, and (b) a deep inelastic $(x,
Q^2)$ + identified forward $\pi^{0}$ $(x_\pi, k_{\pi T})$
event.

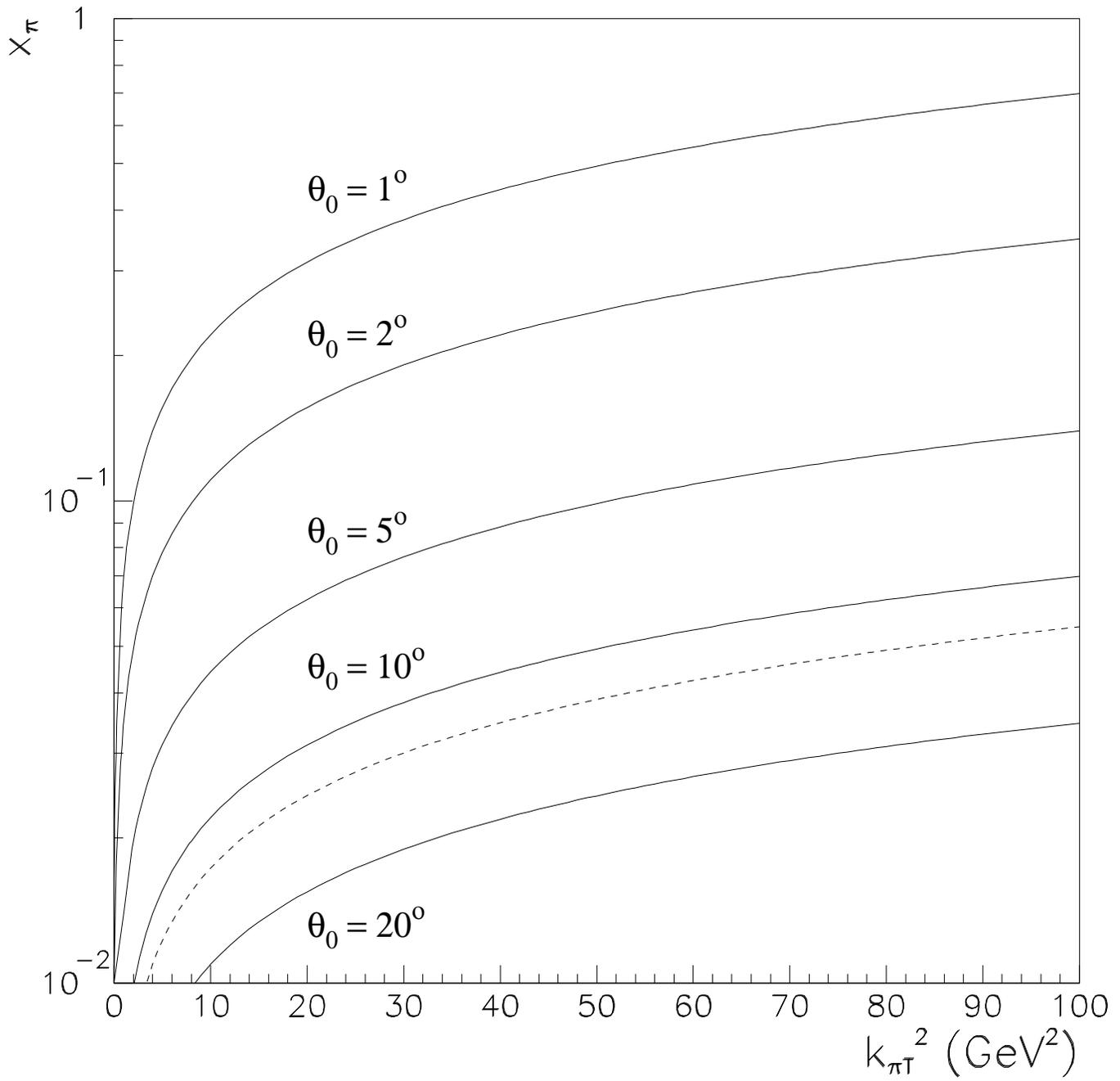
\item[Fig.\ 2] The relation between the $\pi^{0}$ kinematic variables
for DIS + $\pi^{0}$ events with $x = 6 \times 10^{-4}$ and $Q^2 =
20$ GeV$^2$ for various choices of the angle $\theta_{0}$ 
in (\ref{eq:a20}). In the HERA $(27.6 \times
820$ GeV) laboratory frame the pion angle $\theta_{\pi p}$ to 
the proton direction is not uniquely
specified by $(x, Q^2; x_\pi, k_{\pi T}^2)$. Varying the remaining
azimuthal angle transforms the lines of constant $\theta_{\pi p}$
into narrow bands in the $x_{\pi}$, $k_{\pi T}^2$ plane. Here we
averaged over the azimuthal degree of freedom. The plot is 
insensitive to variations of $x$, $Q^2$ over their relevant intervals.
The continuous lines are the {\it upper} boundary on the
allowed kinematic region for different choices of $\theta_0$.  The dashed 
line represents the {\it lower} boundary given by the hemisphere cut, 
(\ref{eq:a17}), for $x = 6 \times 10^{-4}$ and $Q^2 = 20$ GeV$^2$.

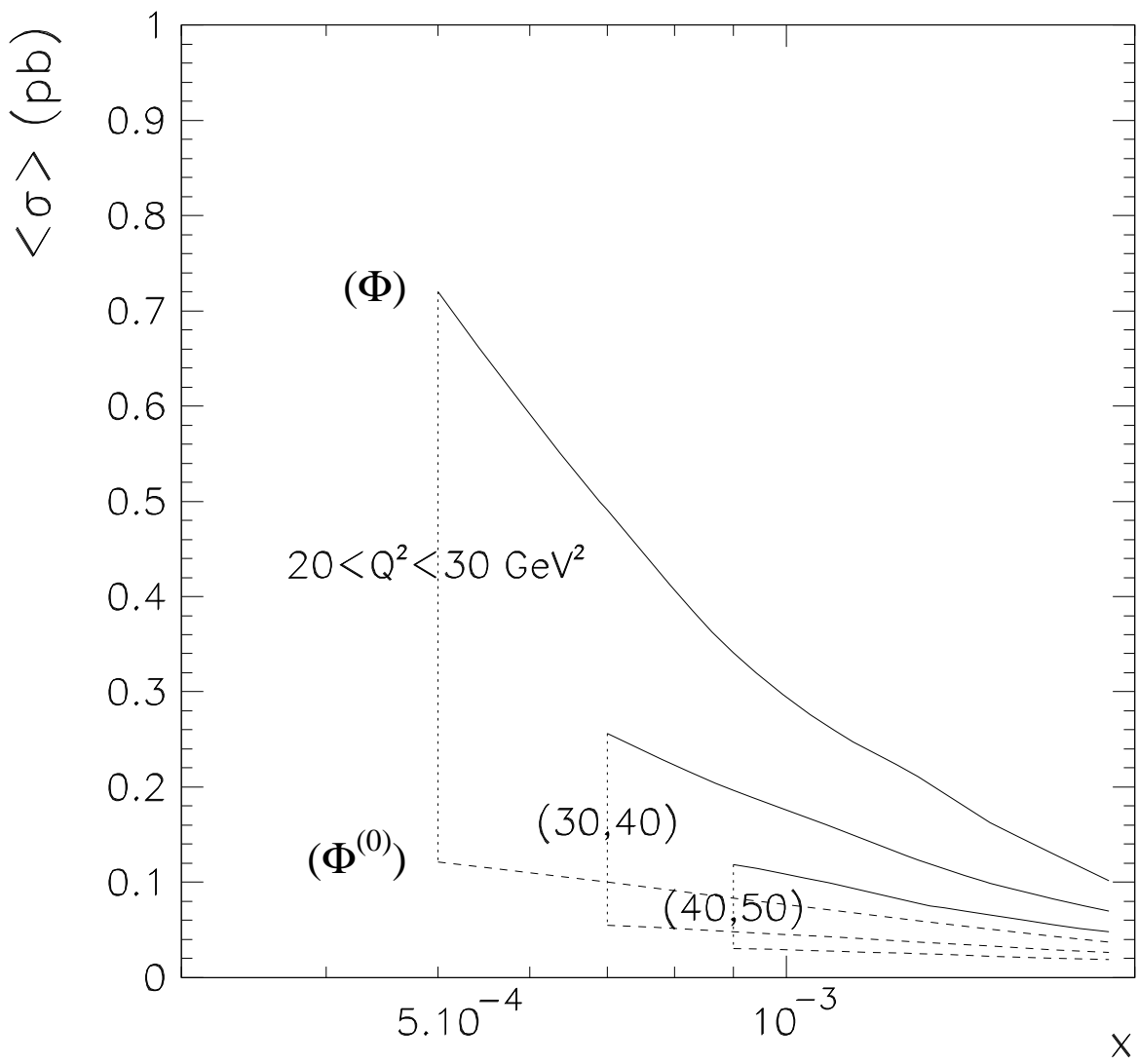
\item[Fig.\ 3] The cross section, $\langle \sigma \rangle$ in pb,
for deep inelastic + $\pi^{0}$ events integrated over bins of
size $\Delta x = 2 \times 10^{-4}$, $\Delta Q^2 = 10$ GeV$^2$
which are accessible at HERA for $\pi^{0}$'s with transverse
momentum $3 < k_{\pi T} < 10$ GeV where the 
constaints $x_{\pi} > 0.05$, 
$\theta_{\pi p} > 5^\circ$, and the hemisphere cut, (\ref{eq:a17}),
were imposed. The fragmentation functions were evaluated at
scale  $k_{\pi T}^{2}$.
The $\langle \sigma \rangle$ values are plotted at 
the central $x$ value in each $\Delta x$ bin and joined by 
straight lines. The $x$ dependence is plotted for three different
$\Delta Q^2$ bins, namely (20,30), (30,40) and (40,50) GeV$^2$.
The continuous curves show $\langle \sigma \rangle$ calculated
with $\Phi_i$ obtained from the BFKL equation. The corresponding
$\langle \sigma \rangle$ values calculated neglecting soft gluon 
resummation and just using the quark box approximation
$\Phi_{i} = \Phi_{i}^{(0)}$ are plotted as dashed curves.
For clarity a dotted vertical lines joins each pair of curves
belonging to the same $\Delta Q^2$ bin.

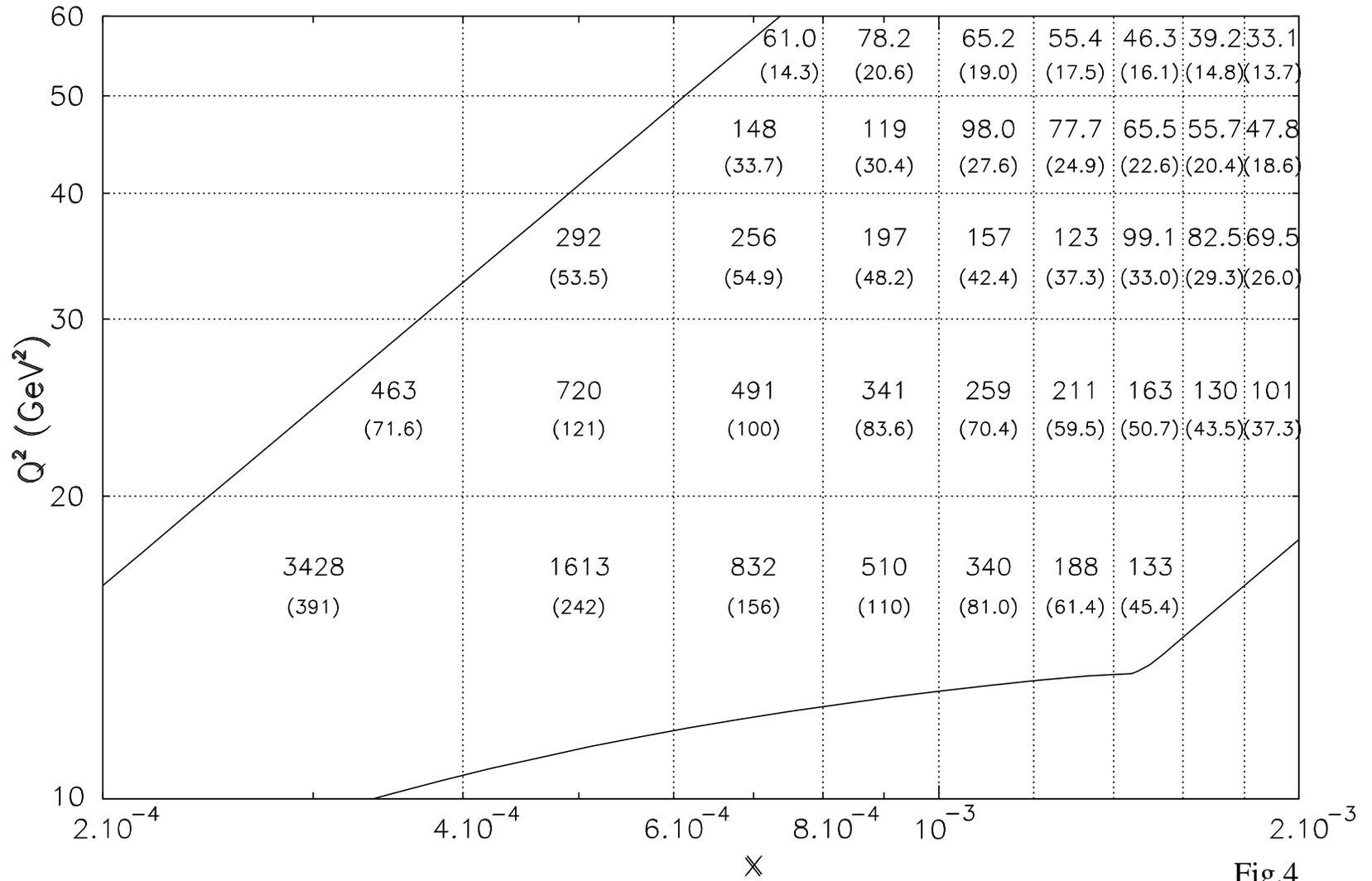
\item[Fig.\ 4] The cross section, $\langle \sigma \rangle$ in fb,
for deep inelastic + $\pi^{0}$ events in various $(\Delta x, \Delta
Q^2)$ bins which are accessible at HERA, and integrated over the
region $3 < k_{\pi T} < 10$ GeV, $\theta_{\pi p} > 5^\circ$, 
$x_{\pi} > 0.05$, and subject to the hemisphere cut, (\ref{eq:a17}).
The fragmentation functions were evaluated at scale  $k_{\pi T}^{2}$.
The values in brackets are the cross sections obtained when using
only the quark box approximation $\Phi_{i} = \Phi_{i}^{(0)}$.
Therefore the difference between the two numbers shown in one
bin is the enhancement due to BFKL soft gluon resummation.
Recall that (\ref{eq:a15}) implies that the results shown
for the DIS + $\pi^{0}$ cross section here equal the average
of the cross sections for $\pi^+$ and $\pi^-$ production.  The curves 
are the boundaries of the acceptance regions at HERA given by
$8^o < \theta_e < 172^o$, $E_e > 5$ GeV and $0.1 < y < 0.9$.

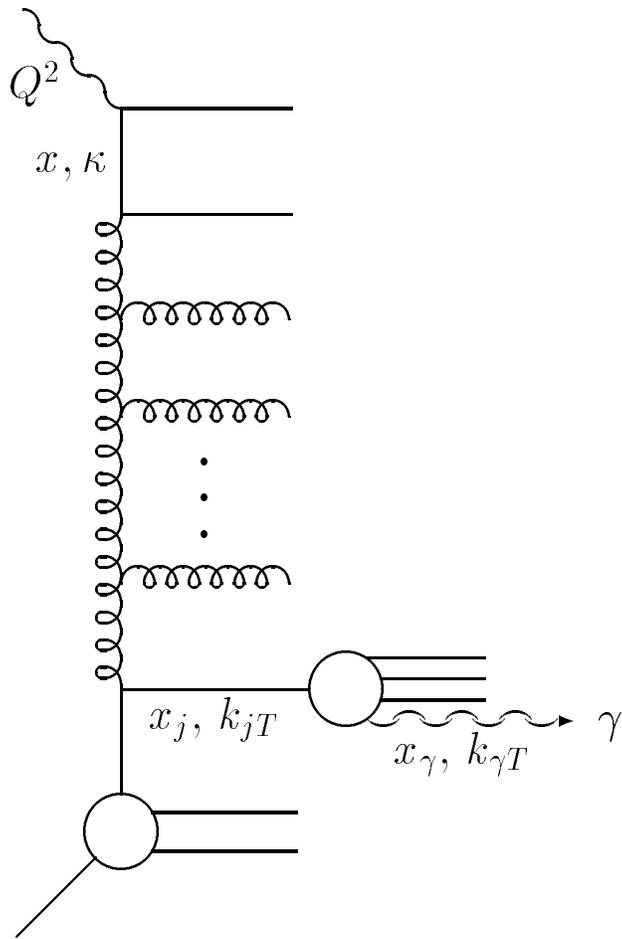
\item[Fig.\ 5] Diagrammatic representation of the background to
deep-inelastic + $\pi^{0}$ events arising from photons which
are fragments of the forward jet.

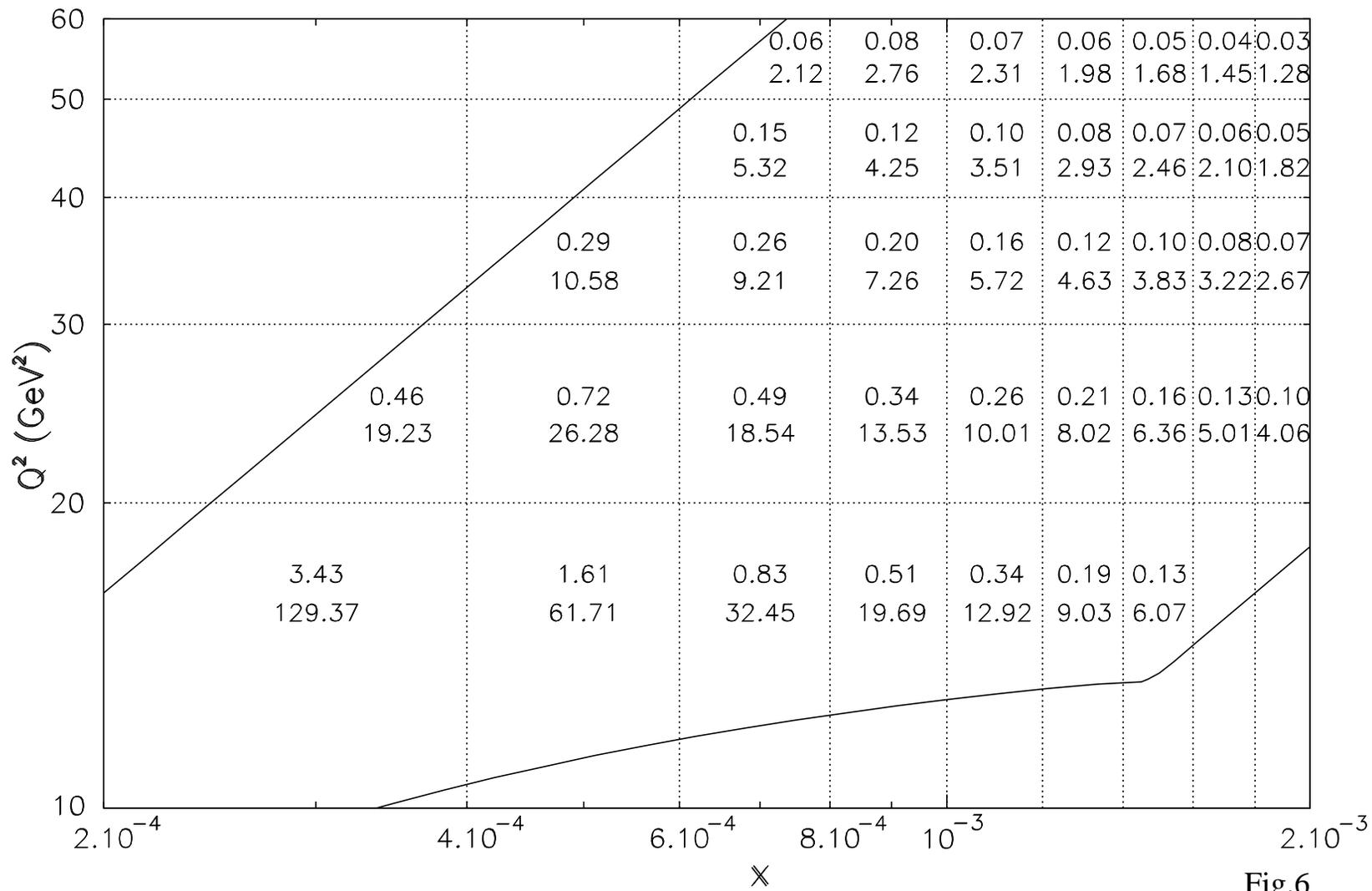
\item[Fig.\ 6] The upper and lower numbers are respectively the 
DIS + $\pi^0$ and DIS + jet cross sections (in pb) in various bins 
of $x$ and $Q^2$.  For the pion the cuts are those given in Fig.~4, 
and exactly the same cuts are used for the forward jet.  

\end{itemize} 

\end{document}